\documentclass{eas}
\usepackage{graphicx}

\newcommand{\Zsun}{{Z_\odot}}
\newcommand{\mdot}{{\dot{M}}}

\begin{document}

\title{Massive close binaries} 
\author{Dany Vanbeveren}\address{Astrophysical Institute, Vrije Universiteit Brussel, Pleinlaan 2, 1050 Brussels}
\begin{abstract}
In the present review we summarize direct and indirect evidence that the massive close binary frequency is very large. We
then discuss the binary evolutionary processes and we present a general massive close binary evolutionary scheme. Finally,
we highlight the importance of massive close binaries for population number synthesis and the chemical evolution of
galaxies.   
\end{abstract}

\maketitle

\section{Introduction}
Since the evolution of a star depends critically on whether or not it is a component of an interacting binary, at 
least for population studies it is essential to know how many stars are binary components. In section 2 I will
first discuss direct and indirect evidence that a very significant fraction of all massive stars are binary
members. The various physical processes acting in binaries will be summarised in section 3. An extended discussion
of massive close binaries and their evolution is given in the monograph {\it The Brightest Binaries}
(Vanbeveren et. al. 1998a); a summary of the latter is given by Vanbeveren et al. (1998b). 

In the following we will use a number of abbreviations which are listed in table 1 below. The primary is defined as 
the binary component which was originally (on the ZAMS) the most massive one.  The secondary is then obviously the other
component. During the RLOF/mass transfer phase, it has also become practice to use the term {\it mass loser} and {\it mass
gainer}. The mass ratio q of a binary equals the mass of the secondary divided by the mass of the primary.

\begin{table}[ht] 
\caption{Abbreviations used in the present paper}
\begin{center}
\begin{tabular}{l@{\hspace{10mm}}l@{\hspace{20mm}}l}

{\small CHB}&{\small Core Hydrogen Burning}\\
{\small HSB}&{\small	Hydrogen Shell Burning}\\
{\small CHeB}&{\small		Core Helium Burning}\\
{\small HeSB}&{\small		Helium Shell Burning}\\
{\small MCB}&{\small	 	Massive Close Binary}\\
{\small SN}&{\small				Supernova}\\
{\small RLOF}&{\small		Roche Lobe Overflow}\\
{\small BH}&{\small				Black Hole}\\
{\small NS}&{\small				Neutron Star}\\
{\small NSM}&{\small			Neutron Star Merger, either two NS star or a BH with a NS}\\
{\small cc}&{\small				compact companion, a NS or a BH}\\
{\small HMXB}&{\small		High Mass X-ray Binary}\\
{\small LMXB}&{\small		Low Mass X-ray Binary}\\
{\small LBV}&{\small			Luminous Blue Variable}\\
{\small PNS}&{\small 		Population Number Synthesis}\\
{\small RSG}&{\small		Red Supergiant}\\
{\small SW}&{\small		Stellar Wind}\\
{\small TZO}&{\small		Thorne-Zytkow Object}\\
{\small OBN}&{\small OB-type stars with anomalous nitrogen lines}\\
{\small OBC}&{\small OB-type stars with anomalous carbon lines}\\
{\small WR}&{\small		Wolf-Rayet}\\ 
{\small WN}&{\small		WR of the Nitrogen sequence}\\
{\small WC}&{\small		WR of the Carbon sequence}\\
{\small ZAMS}&{\small		Zero Age Main Sequence}\\
{\small CE}&{\small		Common Envelope}\\
{\small SpI}&{\small		Spiral-In}\\
{\small SS}&{\small		Single Star}\\
{\small SPH}&{\small		Smooth Particle Hydrodynamics}\\
{\small Min(LBV)}&{\small	The Minimum ZAMS mass of LBVs}\\ 
{\small Min(WR)}&{\small	The Minimum ZAMS mass of WR stars}\\
{\small Min(NS)}&{\small	The Minimum ZAMS mass of NSs}\\
{\small Min(BH)}&{\small	The Minimum ZAMS mass of BHs}\\
{\small WD}&{\small		White Dwarf}\\
{\small CEM}&{\small		Chemical Evolutionary Model (of galaxies)}\cr

\end{tabular}
\end{center}
\end{table}

\section{The massive close binary frequency}

The binary frequency may not be confused with the number of stars in binaries. Suppose that we have 3 O-type stars,
1 single and the two others in 1 binary. Then the binary frequency is 50\% but 2/3 of the O-type stars are binary
members. Furthermore, the observed binary frequency is not necessarily the same as the binary frequency at birth (on
the ZAMS). Stars which are born as binary components may become single stars during the evolution of the binary (due
to the merger process or due to the SN explosion of the original companion, see also section 3). Any observed star
sample contains the binaries, the single stars which are born as single and the single stars which became single but
which were originally binary members. This means that the observed binary frequency is always a lower limit of the
binary frequency at birth.

\subsection{Direct evidence}

The observationally confirmed O-type spectroscopic (thus close) binary frequency in young open clusters has been reviewed
by Mermilliod (2001).  The rates are as high as 80\% (IC 1805 and NGC 6231), and can be as low as 14\% (Trumpler 14). The
MCB frequency in Tr 16 is at least 50\%  (Levato et al., 1991) and in the association Sco OB2 it is at least 74\%
(Verschueren et al., 1996). The results listed above can be considered as strong evidence that the MCB frequency may vary
among open clusters and associations. 

Mason et al. (1998) investigated the bright Galactic O-type stars and concluded that the
observed spectroscopic binary frequency in clusters and associations is between 34\% and 61\%, among
field stars between 20\% and 50\% and among the O-type runaways between 5\% and 26\% (a runaway is defined as a star
with a peculiar space velocity larger than 30 km/s). However, as argued by the authors, these percentages may
underestimate the true MCB frequency due to selection effects. The close systems that are missing in most
of the observational samples are systems with periods larger than 40 days and mass ratios smaller than 0.4 which are
obviously harder to detect. Mason et al. conclude that it cannot be excluded that we are still missing about half of
all the binaries. Accounting for the observed percentages given above, we therefore conclude that the O-type MCB
frequency may be very large. 

Vanbeveren et al. (1998a, b) investigated the bright B0-B3 stars (stars with a mass
between 8 M$_{\odot}$ and  20 M$_{\odot}$) in the Galaxy and concluded that at least 32\% are a primary of
an interacting close binary. Again due to observational selection, using similar arguments as for the O-type
binaries, the real B0-B3 binary frequency could be at least a factor 2 larger.  

Let us remark that the observations discussed above give us a hint about the MCB frequency in the solar
neighbourhood. However, whether or not this frequency is universal is a matter of faith.

\subsection{Indirect evidence}

Indirect evidence about the binary frequency comes from PNS studies. Using all we know about single star and binary 
evolution, we can calculate the number of binaries of some type and compare it with the observed number (see also
section 4). It is obvious to realise that predicted numbers depend on the adopted initial binary frequency; it can
be concluded that in order to obtain general correspondence with observed numbers, the initial MCB frequency in the
simulations must be very large ($>$ 50\%). Interestingly, De Donder and Vanbeveren (2002, 2003b) compared observed
and theoretically predicted SN type Ia rates with PNS of intermediate mass single stars and binaries, and also in
this case the adopted initial intermediate mass close binary frequency must be very large in order to obtain
correspondence.

\subsection{Overall Conclusion}

The MCB frequency could be very large. This means that\\ 

\noindent {\it the conclusions resulting from the interpretation of data
of populations of stars and/or stellar phenomena where the effects of MCBs are denied, may have an academic value but
may be far from reality}.

\section{Massive close binary evolution}

The evolution of a binary component depends on its initial mass, its initial chemical composition, its rotation, on 
the binary period, on the mass of its companion, on the initial deviation from synchronism (the difference between
the rotation period and the orbital period) and on the deviation from circularisation (the eccentricity).  In the
present paper I will discuss general properties of binary evolution which allow to study global effects of binaries
on the evolution of large populations/galaxies.

\subsection{Synchronisation and circularisation}

When the orbit is eccentric and/or the rotation period of a binary component differs from the orbital period, in
addition to rotational mixing (similar as in rotating single stars) a binary component may be subject to tidal
mixing. Since after its formation a star rapidly reaches a state of central mass concentration (including the
stellar core), it is easy to understand that tidal mixing will primarily affect the outer layers of a star and may
add to rotational mixing in order to bring CNO matter to the surface. However the overall evolution of the star is
expected to depend only marginally on the process. Even more, one of the main effects of the RLOF is the removal of
all the outer layers, so that it is fair to neglect the effects of tidal mixing in order to study general properties
of binary evolution.

\subsection{The Roche lobe overflow}

If a massive star in a binary expands, its may fill its Roche volume at a certain moment and RLOF mass loss/mass 
transfer will start. The latter will go on as long as the star keeps its tendency to expand.
 
We distinguish three major expansion phases (see also the contribution of A. Maeder in the present volume): during 
CHB where the evolutionary timescale is the nuclear timescale, during HSB and during HeSB where the star evolves on
the corresponding thermal timescale. When the MCB period is of the order of a few days, the primary will reach its
critical Roche lobe while it is still in its CHB phase: using the terminology of Kippenhahn and Weigert (1967) and
of Lautherborn (1969) this binary will be classified as case A. When the binary has a period between a few days and
a few hundred to thousand days, the primary will fill its Roche lobe while it is a HSB star $=>$ case B. When the
period is larger than 1000 days (up to a few 1000 days), the primary will start its RLOF while it is a HeSB star $=>$
case C. Case B binaries can be further subdivided. During the first part of HSB a massive star has a (mainly)
radiative envelope. In a case B$_{r}$ binary the primary reaches its Roche lobe during this first part. When the
binary period is larger, the primary may reach the RSG phase and most of the outer layers may become convective
before the onset of the RLOF. A case B binary where this can happen is classified as case B$_{c}$. 

\subsubsection{Case A/B$_{r}$}

\noindent {\it case A}: after an initial rapid but short mass loss phase (10$^{-4}$ M$_{\odot}$/yr are typical mass loss
rates), the mass loser continues losing mass due to RLOF on the nuclear timescale (10$^{-6}$ M$_{\odot}$/yr are typical
mass loss rates). At the end of CHB, the mass loser contracts and the RLOF phase stops. However, the star still
contains a considerable part of its original hydrogen rich layers and this means that during the subsequent HSB phase,
the star will expand again and a case B$_r$ type of RLOF will start, i.e. {\it a case A is almost always followed by a case
B$_{r}$}.\\

\noindent {\it case B$_{r}$}: the RLOF typically lasts 10$^4$-10$^5$ yrs and the star may lose mass at a rate
$\approx$ 10$^{-3}$ M$_{\odot}$/yr. The expansion of the star stops when most of the hydrogen rich layers are removed
and when He starts burning in the core. The final remnant mass equals the mass of the CHB
core when the central hydrgen abundance $\approx$ 0.2-0.3. During most of the ensuing CHeB, the star is a hydrogen
deficient star which resembles a WR star, and stays well inside its Roche lobe. After CHeB, during HeSB, the star may
expand again and a second RLOF phase starts. The current name of this phase is case BB RLOF (section 3.2.3).\\ 

During a case B$_r$ RLOF phase, the mass transfer rates can be very large and the mass gainer may expand
significantly  (see subsection 3.3). Using the most up to date opacity tables (Iglesias and Rogers, 1996), it
follows that almost all massive case B$_r$ binaries will evolve into contact (both components fill a common
equipotential surface). Depending on the initial binary parameters, a situation can occur where matter lost by the
loser escapes the binary. This is always accompanied by a very large orbital angular momentum
loss (Soberman et al., 1997) implying a significant reduction of the orbital separation.

\subsubsection{Case B$_c$/C: the common envelope evolution}

Mass loss due to RLOF in a case B$_c$ or case C binary happens on the dynamical timescale. The gainer is unable to 
accrete the mass lost by the loser at the same rate and soon after the onset of the RLOF a CE is formed. The physics
of CE evolution has been reviewed by Iben and Livio (1993). Among the updates later on we like to promote the study
of Rasio \& Livio (1996) who used the smoothed particle hydrodynamics (SPH, Monaghan, 1992; Rasio \& Shapiro, 1991,
1992) in order to study the CE phase of low and intermediate mass binaries where the masses of the two stars are
very different. The CE evolution of a binary where both stars have very similar masses and/or both stars are massive
has not yet been studied in detail but based on qualitative arguments and the computations listed above, a very
plausible model is the following:\\

\noindent {\it viscous drag dissipates orbital energy into the CE; part of it is radiated away, part is used to expel
the CE; one of the main effects is a large reduction of the binary separation and the possible merging of the two
stars.}\\

Detailed numerical simulations reveal that this CE/SpI is very rapid. Therefore, if one is mainly interested in the 
post-CE remnant rather than in the details of the process itself, one may use the energy prescription as proposed by
Webbink (1984).  Depending on what happens with this common envelope, the mass loss of the loser stops when most of
the hydrogen rich layers are removed (remnant mass similar as in case B$_r$) or when the two components merge. When
merging is avoided, also case B$_c$ may be followed by case BB.

\subsubsection{ Case BB} 

The RLOF in a binary stops when helium starts burning in the core of the mass loser and when most of the hydrogen 
rich layers have been removed: the mass loser has become a hydrogen deficient CHeB star. When the post-RLOF mass is
smaller than 5 M$_{\odot}$ (corresponding with an initial mass on the ZAMS smaller than 15  M$_{\odot}$) the further
evolution deserves some attention. Habets (1986a,b) computed the evolution of helium stars with 2 $\le$ M/
M$_{\odot}$ $\le$ 4 up to neon ignition and concluded that those with 2 $\le$ M/ M$_{\odot}$ $\le$ 2.9 develop deep
convective envelopes during the HeSB phase, after CHeB, and expand significantly. Depending on the binary period,
these stars may fill their Roche lobe again and perform case BB RLOF. During this phase of mass transfer the star
loses its remaining hydrogen layers and most of its helium layers on top of the He burning shell. As has been
outlined in Vanbeveren et al. (1998a), one of the most important effects of case BB RLOF in relation to population
synthesis is the orbital period evolution of the binary. When the companion is a normal mass gainer, case BB RLOF
results into a very large period increase. When the companion is a compact star, case BB RLOF is governed by the
SpI process (subsection 3.4) which may result in a significant hardening of the binary. The latter has a large
effect on the probability for the binary to remain bound after the second SN explosion, and thus on the birth rate
of double compact star systems (subsection 4.4). 

However, massive post-RLOF CHeB stars may lose mass by stellar wind. Vanbeveren et al. (1998a,b)
argued that when the stellar wind mass loss rate formalism of WR stars (which have a mass larger than 5  M$_{\odot}$) is
extrapolated downwards, it cannot be excluded that this mass loss is sufficiently large in order to suppress case BB
RLOF in massive binaries. With this SW scenario the orbital period hardly varies. As far as PNS of double compact
star binaries is concerned, this is probably the most important difference compared to the case BB RLOF scenario.
Notice that when the WR mass loss rates scale with the iron abundance according to Equation~\ref{eq:mom} (subsection 3.8),
case BB RLOF will not be suppressed in low Z environments.

\subsection{ The evolution of the mass gainer}

The evolution of the mass loser in a binary is reasonably well known. The evolution of the mass gainer however 
is very uncertain. The real situation is the following: a gasstream leaving the loser through the first Lagrangian
point is attracted by the gainer. Depending on the binary separation, this gasstream either hits the gainer directly
or it first forms a Keplerian disc around the gainer. Accretion of matter is accompanied by the accretion of angular
momentum and the outer layers of the star may spin up. How this spin up process is further transported inward is
unknown, however, when a Keplerian disc can be formed, it is easy to demonstrate that there is enough angular
momentum available in the disc in order to spin up the whole star up to a rotational velocity which is close to the
critical value. To compute the real consequences of mass accretion during RLOF would require the solution of the
full magneto-hydrodynamic equations of the gasstream and a theory in order to simulate the reaction of a star when
this gasstream hits its surface: approximations are necessary. The following scenarioÕs can be considered as
limiting cases.\\

\noindent {\it Scenario 1} (Neo et al., 1977): accretion is treated in a spherically symmetric way; the matter has a
specific entropy which is larger than or equal to the one of the outer layers and spin-up is neglected. The chemical
composition of the gasstream and of the outer layers of the gainer may be different and this may initiate
thermohaline mixing (Kippenhahn et al., 1980).\\

\noindent {\it Scenario 2} (Vanbeveren and De Loore, 1994): accretion of mass implies accretion of angular momentum
and large rotation may induce efficient mixing. The limiting situation is of course a complete mixing of the
whole star.\\

Detailed calculations allow to propose the following conclusions.

\begin{itemize}

\item If, during mass transfer, scenario 1 applies, the mass gainers in all case B MCBs expand and most (all) of them
reach their own Roche limit during the RLOF phase of the loser, i.e. the components in the majority (all) of the
case B MCBs evolve into contact. 
\item If, during mass transfer, scenario 2 applies, a large expansion and contact can
be avoided. 
\item Mass gainers will be observed as nitrogen enriched stars (classified as OBN); this nitrogen enrichment
is more pronounced when scenario 2 applies. 
\item Mass gainers may show up as rejuvenated stars, i.e. they will look
younger than their accompanying mass loser. In a starburst, they will be observed as blue stragglers. 
\item When scenario 1 applies and the binary has an initial mass ratio close to one, depending on the treatment of
convection inside a massive star, the mass gainer may remain in the blue part of the HR diagram during its remaining
life and an excursion towards the red (which is typical for normal massive single stars) will not happen; the latter
scenario may explain the blue nature of the progenitor of SN 1987A (De Loore \& Vanbeveren, 1992; Braun \& Langer,
1995).

\end{itemize}

\subsection{The spiral-in phase}

Binaries where the mass ratio $\le$ 0.1 meet the Darwin instability (Sparks \& Stecher, 1974). Instead of a classical
 RLOF, the low mass star is engulfed by the higher mass component.

The evolutionary computations of case B$_r$ MCBs indicate that although a mass loser shrinks in response to mass
loss, it may not be able to shrink fast enough to keep up with the rapid shrinkage of the Roche lobe when initially
the binary has a mass ratio 0.1 $<$ q $\le$ 0.2. Also in this case, the low mass component will be dragged into the
envelope of its higher mass companion. 

It is therefore conceivable that in most of the binaries with initial q $\le$
0.2, the low mass star will be engulfed by its higher mass companion. The further evolution will then be governed by
viscous forces and the low mass star will spiral-in.  SpI will stop when the two stars merge or when there is
sufficient orbital energy available in order to expel the hydrogen rich layers of the massive star. If one is mainly
interested in the remnant after SpI, a formalism can be used similar as the CE formalism (section 3.2.2).

\subsection{Mergers}

The SpI in binaries with q $\le$ 0.2, a non-conservative RLOF during contact or the CE in case B$_c$/C binaries may
lead to the coalescence of both stars: the stars merge. Mergers may be very important in overall PNS but the physics
of merging and how mergers evolve afterwards are very uncertain. SPH (Monaghan, 1992; Rasio \& Shapiro, 1991, 1992)
may be very useful to study them, especially SPH in combination with stellar evolutionary codes. However, a
systematic study has not been done yet. 

When both stars are non-degenerate, merging could be considered as an
extreme case of mass and angular momentum accretion and therefore the evolution of mergers will most probably be
influenced by rotational diffusion. Similarly as with accretion of mass during the RLOF, a possible outcome of
merging may be a star that is or will be mixed and will be composed of material from two stars which have altered
CNO abundanceÕs or, more spectacular, which may contain the products of the 3$\alpha$ process. In the latter case,
the resulting star may be carbon enhanced and could be classified as OBC. 

When the low mass companion is a degenerate star, merging means that spiral-in continues until the compact star
reaches the centre of the star: a TZO is formed (Thorne \& Zytkow, 1977). Evolutionary calculations of Cannon et
al. (1992) reveal that a TZO resembles a RSG. The further evolution is governed by SW mass loss; when all the mass is
lost the compact star emerges again. 

A special merging process happens in binaries which consists of two NSs or of a BH and a NS. Two compact stars
spiral together as a consequence of gravitational wave radiation. A formalism to calculate typical decay
timescales as function of the initial orbital parameters of the double compact star system has been worked out by
Peters (1964).  Gamma-ray bursts may be related to the merging of both stars and they may be an important
production site of r-process elements (see section 5).

\subsection{The effect of the SN explosion on a binary} 

It has become clear the last decade that the SN explosion of a massive star is not spherically symmetric. The 
importance for binary evolution (binary remains bound or is disrupted) is obvious if one accounts for the fact that
only a small asymmetry is sufficient in order to impart a very large kick velocity to the compact remnant. The
effects of an asymmetric SN explosion on binary orbital parameters has been worked out in detail by Tauris and
Takens (1998). Relating the observed space velocities of single pulsars (Lorimer et al., 1997) to the kick velocity,
detailed simulations reveal that in most of the cases a massive binary is disrupted as a consequence of the SN
explosion of one of its components. 

A double NS system has survived two SN explosion. This means that although the number of observed double NS systems is
small, the fact that there are a few means that the progenitor binary frequency must be large (this is sustained by
detailed PNS calculations). 

An important effect of the SN explosion of a star in a binary is the $\alpha$-element pollution of the outer layers
of the companion. The $\alpha$-elements in the atmosphere of the optical companion star of the LMXB GRO J1655-40
(Nova Sco 1994) observed by Israelian et al. (1998) strongly support the scenario where the BH formation was
preceded by some SN-like mass ejection. The latter is sustained by the fact that some of the LMXB-BH candidates
and the HMXB Cyg X-1 have large space velocities which may be an indication that SN-like mass ejection occurred
prior to BH formation (Brandt et al., 1995; Fryer and Kalogera, 2001). 

Notice that the pollution due SN explosion in a binary may also explain the existence of OBC stars:
the layers which are expelled during the SN of a massive star may contain a lot of carbon (and oxygen). Part of
these layers hit the companion and simulations reveal that the latter becomes significantly overabundant in carbon
(possibly also in oxygen), i.e. an OBC star forms.

\subsection{Deviations from the previous evolutionary picture}

Deviations from the general RLOF picture summarised above are due to the effects of violent SW mass loss phases. 
Based on the observations of LBVs one may suspect that stars with initial mass $\ge$ 40  M$_{\odot}$ experience an LBV
phase at the end of CHB and/or HSB, during which very high SW mass-loss takes place (Humphreys \& Davidson 1994). The
lack of RSGs with initial mass $\ge$ 40  M$_{\odot}$ (corresponding roughly to stars with M$_{bol}$ $\le$ -9.5, Humphreys
\& McElroy 1984) may be attributed to the LBV phase so that a working hypothesis for stellar evolutionary
calculations may be the following:\\

\noindent {\it the mass loss rate during the LBV + RSG phase of a star with initial mass $>$ 40  M$_{\odot}$  must be
sufficiently large to  assure a RSG phase which is short enough to explain the lack of observed RSGs with M$_{bol}$
$\le$ -9.5.}\\

Obviously, binary components with an initial mass $\ge$ 40  M$_{\odot}$ will obey this criteria as well and this means
that\\

\noindent {\it when a primary with initial mass $\ge$ 40  M$_{\odot}$ starts its LBV phase before the RLOF (case B and case
C systems), the evolution is governed by the LBV stellar wind, and the RLOF (and thus mass-transfer) is suppressed.}\\  

The data of Humphreys \& McElroy (1984) illustrate that also the Magellanic Clouds are deficient in RSGs with
M$_{bol}$ $\le$ -9.5 so that the {\it LBV scenario} may apply in low metallicity regions as well. Notice that the
LBV scenario was introduced more than a decade ago (Vanbeveren, 1991).  

This LBV scenario is particularly important for the evolution of binaries with extreme mass ratio (e.g., a 50  M$_{\odot}$
+ 2  M$_{\odot}$). If the LBV scenario would not apply, SpI would happen and merging would be inevitable, i.e. it would
become very difficult in order to explain the existence of binaries with a massive BH and a low mass companion. However,
the LBV wind reduces the importance of SpI and the binary may survive and become a massive BH + 2  M$_{\odot}$ binary.

Stars with an initial mass $\ge$ 30  M$_{\odot}$ reach the RSG phase and will be subjected to a strong low velocity wind.
When such a star is a component of a case B$_c$ or case C system, it is clear that the CE evolution will be influenced by
this SW mass loss; the CE phase may even be avoided (the ÔRSG scenarioÕ of MCBs, Vanbeveren, 1996).

\subsection{The formation of high mass BH}

Explaining the existence of massive BHs (masses $>$ 10  M$_{\odot}$) is inseparable from a discussion about the SW mass
loss during CHeB (the WR phase). Using a hydrodynamic atmosphere code in which the stellar wind is assumed to be
homogeneous, Hamann et al. (1995) determined  $\mdot$-values for a large number of WR stars. Since then evidence has grown
that these winds consist of clumps and that a homogeneous model overestimates $\mdot$ , typically by a factor 2-4 (Hillier
1996, Moffat 1996, Schmutz 1996, Hamann \& Koesterke, 1998).  Our preferred WR  $\mdot$-formalism that we use already
since 1998, has been discussed in detail in Vanbeveren et al. (1998a, b, c) (see also Van Bever \& Vanbeveren, 2003)
and is the following,

\begin{equation}
Log(-\mdot) = Log L -10 + 0.5Log(X_{Fe}/X_{Fe,solar})
\label{eq:mom}  		
\end{equation}			

\noindent with X$_{Fe}$ the iron abundance. Assuming that the WR SW is radiation driven, we expect that the heavy elements
(primarily iron) are the important wind drivers and thus that $\mdot$  mainly depends on the iron abundance. In Figure 1
we show the resulting pre-SN masses of evolutionary sequences performed with Equation~\ref{eq:mom}. It follows that
Galactic stars with initial mass between 40 M$_{\odot}$ and 100 M$_{\odot}$ end their life with a mass between 10
M$_{\odot}$ and 20 M$_{\odot}$ corresponding to carbon-oxygen (CO)  cores masses between 5 M$_{\odot}$ and 15 M$_{\odot}$.
Figure 1 also shows the final masses of massive stars in the SMC (Z = 0.1$\Zsun$), assuming that
the WR mass-loss rate scales proportional to $\sqrt{X_{Fe}}$ .

\begin{figure}[tb]
\begin{center}
\includegraphics[width=12.5cm]{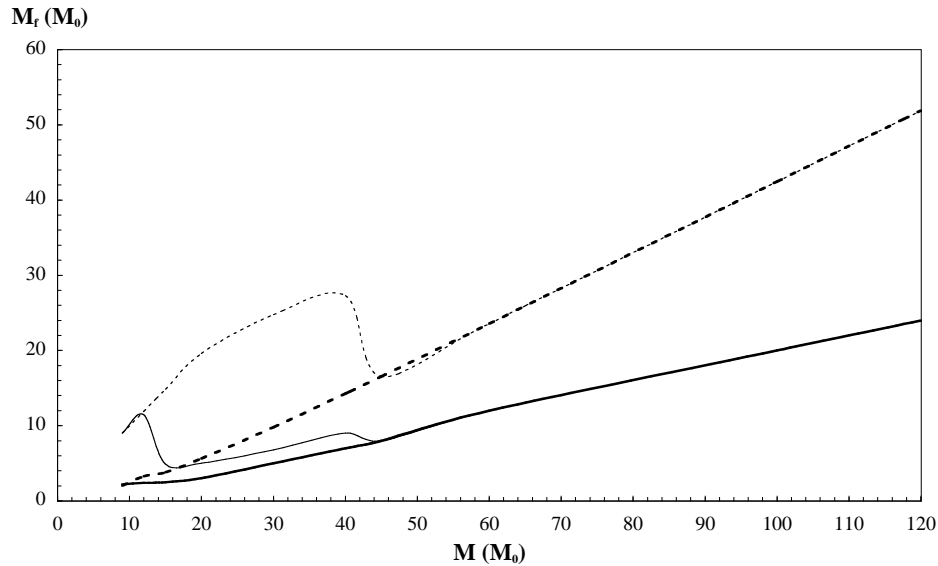}

\caption{The pre-SN masses (M$_{f}$) of massive single (thin lines) and interacting primary (thick lines) stars as predicted by our calculations. The computations
are made for Z=0.02 (full lines) and Z=0.002 (dotted lines).}
\label{fig:fig2}
\end{center}
\end{figure}

\subsection{Rotation, rotational mixing and MCB evolution}

The present volume includes a pleading where the author defends the thesis that rotation and rotational mixing are 
essential ingredients in order to understand massive star evolution. Let me try to discuss the thesis in function of
binaries.  

It is clear that to explain the properties of individual stars (e.g., the CNO surface abundanceÕs are
probably the most important ones), the effects of rotation may be very important. But what about global (average)
evolutionary properties? Compared to non-rotating stellar models, models with rotation have larger convective cores
during CHB. Since the remnant mass after the RLOF/CE phase equals the hydogen deficient CHB core (section 3.2), we can
conclude that the larger the rotation the larger the remnant masses after RLOF. As far as the evolution of primaries in
MCBs is concerned, this is the major effect of rotation (remember that during the RLOF and/or CE phase most of the
hydrogen rich layers leave the loser, layers which may have been affected by rotational diffusion, but who cares in this
case). On average, the effect of rotation on the convective core during CHB is similar to the effect of moderate
convective core overshooting (Vanbeveren, 2001; see also A. Maeder in these proceedings). Since most of the published
evolutionary calculations of MCBs (and also of intermediate mass close binaries) the last decade account for a moderate
amount of convective core overshooting, it can be stated that\\

\noindent {\it On average, MCB evolution the last decade already accounted for the effects of rotation avant la lettre, at
least as far as the  evolution of the loser is concerned.}\\

After the RLOF, the CHeB phase of the remnant is too short for rotational diffusion to play an important role. Notice that 
it can be expected that (again on average) the effect of rotation on SW mass loss during CHeB cannot be very large.
The reason is that in most of the stellar evolutionary codes, semi-empirical SW mass loss rate formalisms are used,
and the effect of rotation on these empirical rates is rather small (Petrenz and Puls, 1996, 2000). 

Notice that rotation and rotational mixing may be very important for the post-RLOF evolution of the mass gainers
in case A/B$_r$ binaries and for mergers (subsections 3.3 and 3.5).

\subsection{The overall MCB evolutionary scenario}

Figures 2-5 summarise the overall MCB evolutionary scenario using the abbreviations listed in the introduction and 
the processes described above.

\begin{figure}[tb]
\begin{center}
\includegraphics[width=12.5cm]{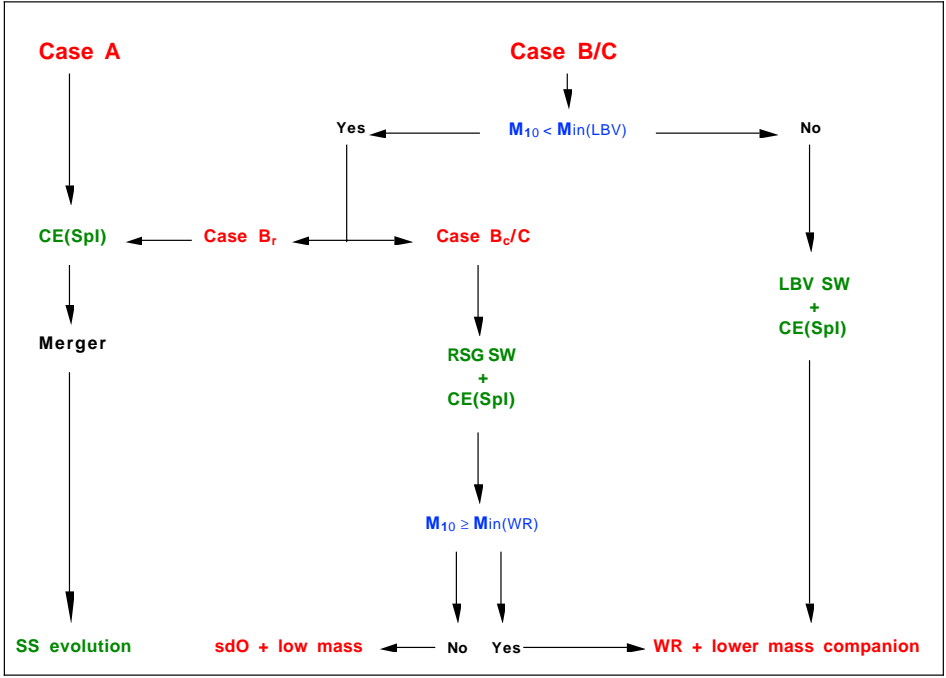}
\end{center}
\caption[]{The evolutionary scenario till the end of RLOF of case A, case B and case C MCBs with initial q $\le$ 0.2}
\label{fig:w1}
\end{figure}

\begin{figure}[tb]
\begin{center}
\includegraphics[width=12.5cm]{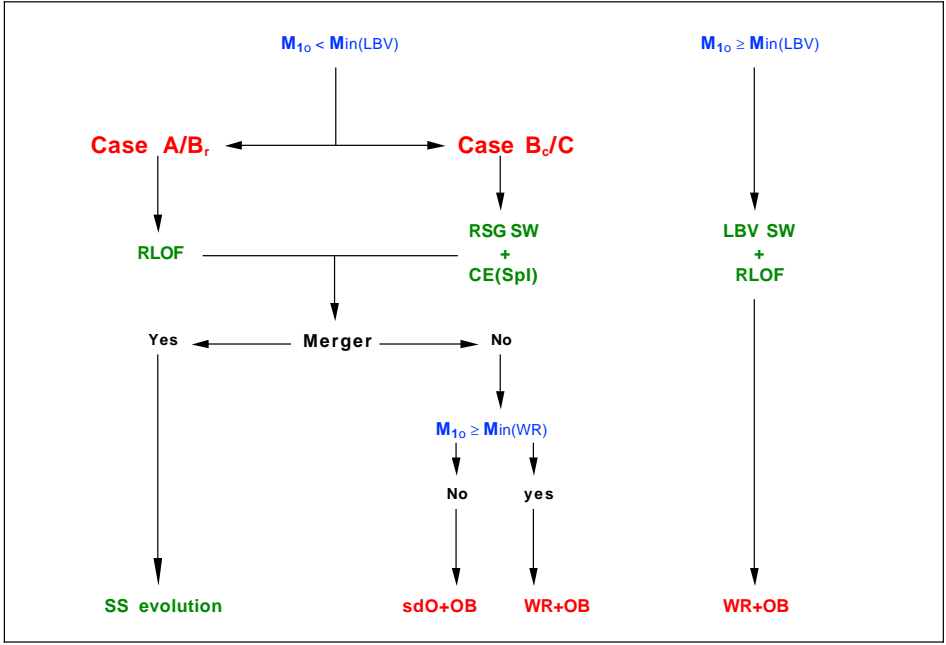}
\end{center}
\caption[]{The same as Figure 2 but for MCBs with initial q $>$ 0.2}
\label{fig:w2}
\end{figure}

\begin{figure}[tb]
\begin{center}
\includegraphics[width=12.5cm]{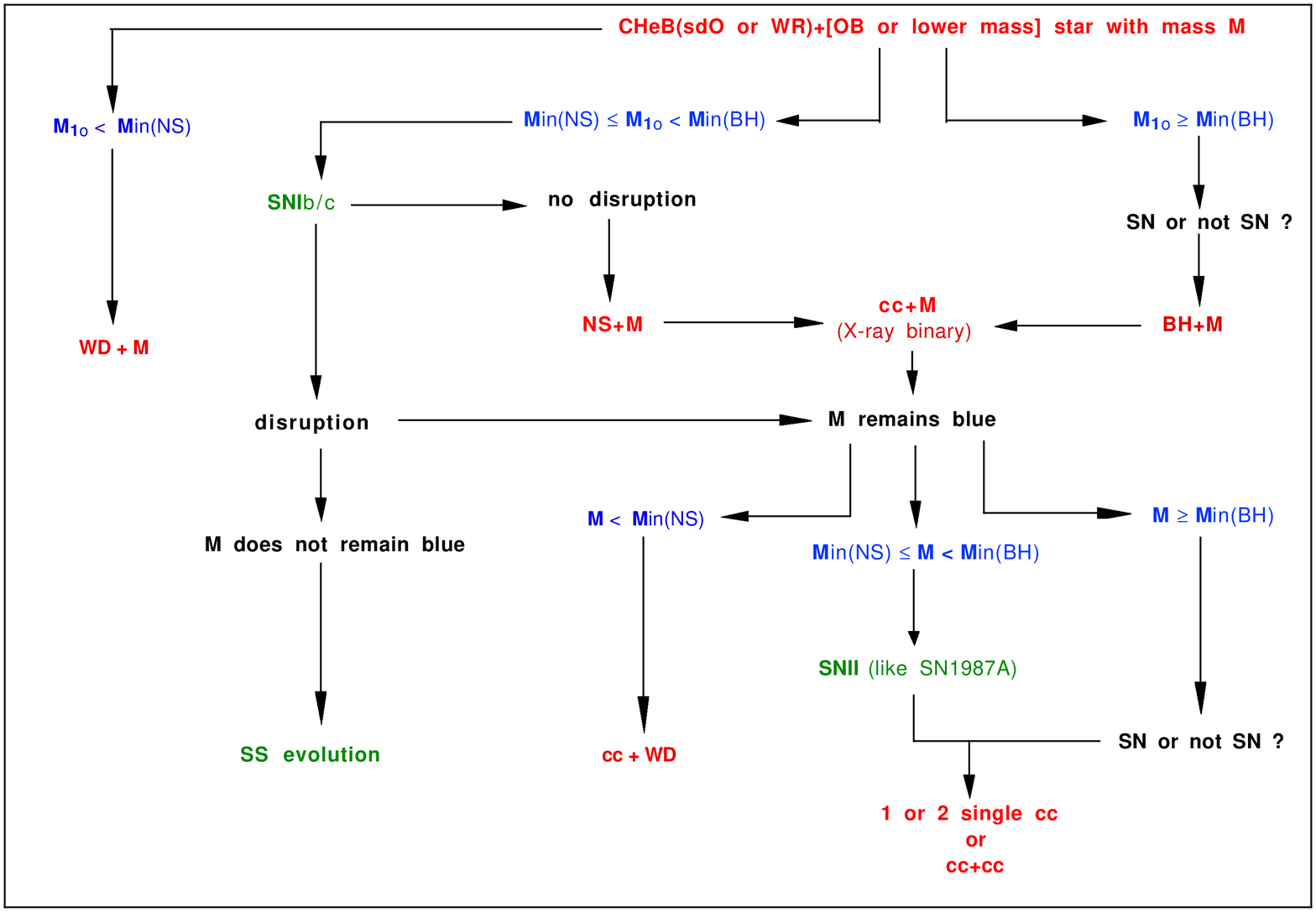}
\end{center}
\caption[]{The post-RLOF evolutionary scenario of MCBs}
\label{fig:w3}
\end{figure}

\begin{figure}[tb]
\begin{center}
\includegraphics[width=12.5cm]{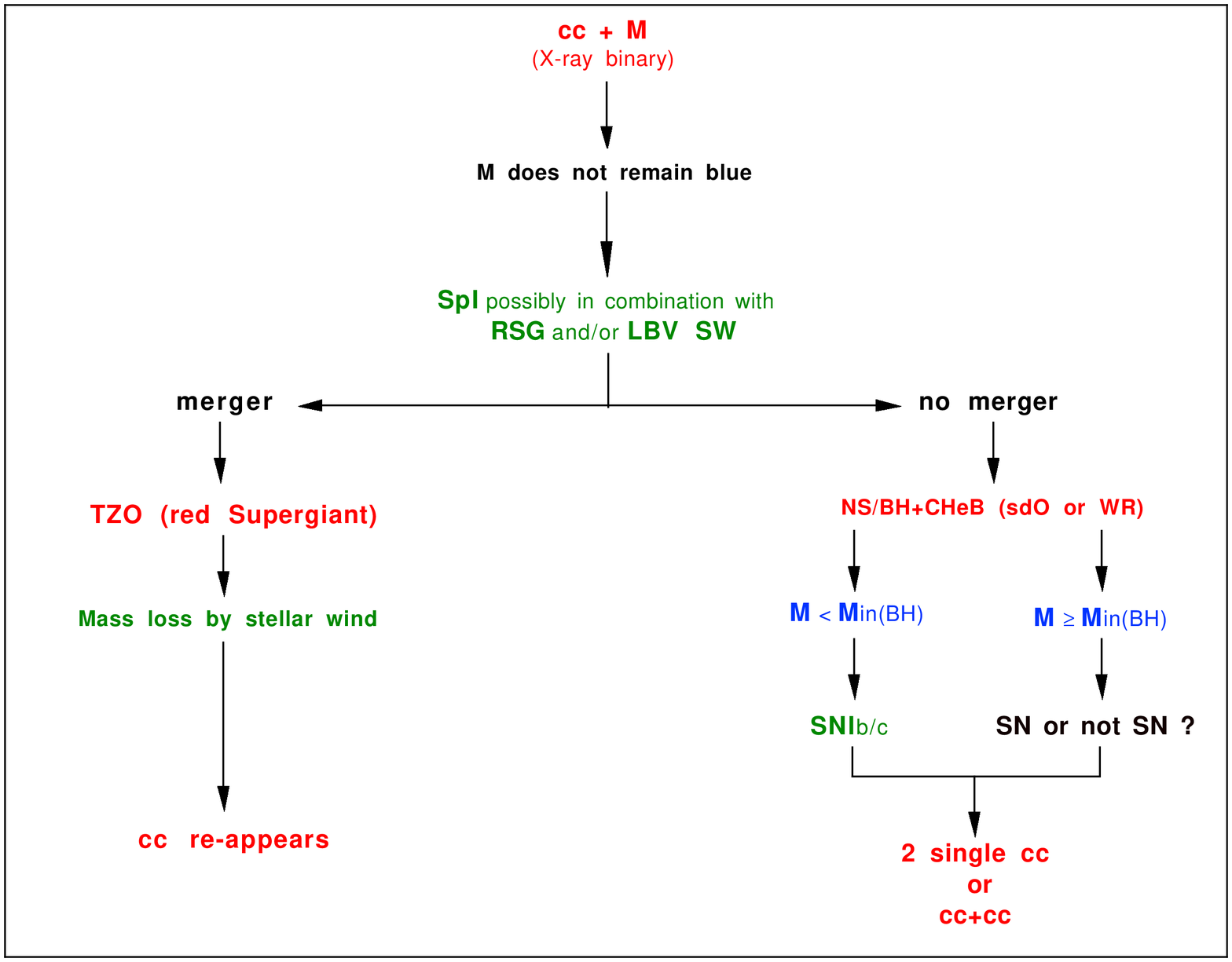}
\end{center}
\caption[]{The post-RLOF evolutionary scenario of MCBs (continued)}
\label{fig:w4}
\end{figure}

\section{Massive star population number synthesis with binaries}

A PNS code with single stars and interacting binaries has to account for:

\begin{itemize}

\item the evolution of single stars
\item the evolution of case A, case B$_r$, case B$_c$ and case C binaries with a mass ratio q $>$ 0.2 
\item the evolution of binaries with a small mass ratio (q $\le$ 0.2) 
\item the evolution of mergers for which we make a distinction between the merging of two normal stars, a normal star
and a compact star and of two compact stars. 
\item the effects of one or two (asymmetric) SN explosions on the binary parameters 
\item a detailed treatment of the evolution of the binary periods, which depends critically on the physics of
the processes summarised in section 3.

\end{itemize}

A PNS code uses distribution functions to initialise the stellar parameters (initial mass function, period and mass
ratio distributions) and to model the asymmetry of a SN explosion.

The Brussels code relies on a large set of stellar evolutionary computations of intermediate mass and massive single stars 
and binary components of which the evolution is followed up to the end of their CHeB phase, with a moderate amount
of convective core overshooting, using the opacity tables of Iglesias \& Rogers (1996) and for the metallicity range
0.001 $\le$ Z $\le$ 0.02. The evolutionary tracks of massive stars are computed with the most recent SW
$\mdot$-formalisms. 

With this PNS skeleton, we studied populations of various stellar types, in
regions of continuous star formation. In particular, we considered the population of Be-type stars (Van Bever \&
Vanbeveren, 1997), the WR and O-type stars (De Donder et al., 1997; Vanbeveren et al., 1998a, b, c), the double
compact star binaries (De Donder \& Vanbeveren, 1998, 2003c), the supernova rates (De Donder \& Vanbeveren, 1998,
2003b). We also adapted our PNS code so that it can follow the stellar populations of starbursts. In particular, we
studied the Be-type stars (Van Bever \& Vanbeveren, 1997), the O and WR population and their effect on the spectral
characteristics of starbursts (Van Bever et al., 1999; Van Bever \& Vanbeveren, 1998; Van Bever \& Vanbeveren, 2003;
Belkus et al., 2003), the X-radiation produced by X-ray binaries and the population of young supernova remnants (Van Bever
\& Vanbeveren, 2000). Of course, there are many groups who studied these topics as well. Many references are given in the
papers listed above. Below, we list a few conclusions.

\subsection{The Be stars}

\begin{itemize}

\item PNS predicts that less than 50\% of the observed Be stars are mass gainers in evolved binaries; the Be components in
Be X-ray binaries obviously belong to this class. Most probably there are many Be stars which are currently observed as
single but have a companion like in $\phi$ Per. 

\item Be stars have initial masses $\le$ 20 M$_{\odot}$ and are found all along the main sequence.

\item The number of B-type star with rotational velocities which are as large as those of Be type stars outnumber the
number of Be type stars; this means that rotation is necessary for the Be phenomenon but it is not sufficient.

\item Significant by their absence is the complete lack of Be + B or Be + Be binaries. Could this be an indication that
rotation in binaries is less important than in single stars?

\item When considering the fraction of Be stars among the corresponding B-type stars, why is there such a large
difference between the Galaxy and the SMC? The Geneva group argued that the effect of metallicity on SW mass loss
and the effect of SW on rotation during CHB could explain this difference. I do not think that this is correct. Be
stars originate from stars with initial mass $\le$ 20 M$_{\odot}$ where the SW is small independent from Z. As a matter
of fact, this is sustained by evolutionary calculations with rotation of the Geneva group itself, i.e. the rotational
velocities in stars with initial mass $<$ 20 M$_{\odot}$ are similar in the Galaxy and in the LMC/SMC.

\item PNS predicts that a significant fraction of all the Be single stars could be mergers resulting from binaries with an
initial mass ratio $\le$ 0.2; this may explain why Be stars are found all along the CHB main sequence.

\end{itemize}

\subsection{The O-type stars, the WR stars and the HMXBs}

\begin{itemize}

\item A significant fraction of the progenitors of WR single stars were binary components. 

\item Very few WR stars are expected to have a cc.

\item The variation as function of metallicity Z of the WC/WN number ratio is indirect evidence that the WR mass loss
rates are Z-dependent. 

\item At least 50\% of the OB-type runaways are formed through the binary SN scenario; since the SN explosion of a binary
component disrupts the binary in most of the cases, only few O-type runaways are expected to hace a cc, corresponding to
observations. The low binary frequency (normal OB-type companion) among O-type runaways (subsection 2.1) is consistent
with the fact that many runaways are formed through the SN scenario.  

\item The temporal evolution of the O-type star and WR star population in starbursts is critically dependent on the
binary population.

\item It is very hard to explain the X-ray data of starbursts without considering the influence of a significant
population of interacting MCBs.

\end{itemize}

\subsection{The SN rates}

\begin{itemize}

\item The predicted SN number ratio Ibc/II depends strongly on the adopted MCB frequency. To explain the cosmic
ratio, an average cosmic MCB frequency arround 50\% seems to be required

\item By comparing the observed and predicted SN number ratio Ibc/II, we conclude that the MCB frequency in late type
spirals may be a factor 2 smaller than in early type spirals.

\end{itemize}

\subsection{The population of double compact star binaries}

\begin{itemize}

\item The theoretically predicted population of double compact star binaries depends critically on whether case BB
RLOF happens or not in MCBs where the mass loser has a pre-RLOF mass $\le$ 15 M$_{\odot}$ (subsection 3.2.3).

\item The binary evolutionary parameters which affect most the predicted properties of the population of double
compact star binaries, are the average kick velocity which describes the asymmetry of the SN explosion and the energy
efficiency parameter during the CE/SpI process of the OB + NS/BH binaries.

\item Our simulations predict a Galactic NS + NS formation rate between (10$^{-6}$ and 10$^{-4}$)/yr.

\item Most of the double NS binaries have a very large eccentricity at the moment of their formation; notice that in
order to calculate the timescale of orbital decay, the knowledge of e is essential (Peters, 1964).

\item Double NS binaries may have very large space velocities; when we combine the orbital decay timescale with these
space velocities, we conclude that merging of the two NSs can happen when the double NS binaries have moved more
than 100 kpc away from their birth place. This means that double NS binaries that are formed in the galactic disk
may be important r-process enrichment sources of the halo, also during the early evolution of the galaxy. 

\item When Z $\le$ 0.002 (representative for the early Galaxy) case BB RLOF is not suppressed by SW during CHeB
(subsection 3.2.3); it follows that most of the double NS binaries merge within a few Myr. We can therefore expect that the
NSM events may be important candidates for the r-process element enrichment in the early Galaxy.

\end{itemize}

\begin{figure}[tb]
\begin{center}
\includegraphics[width=12cm]{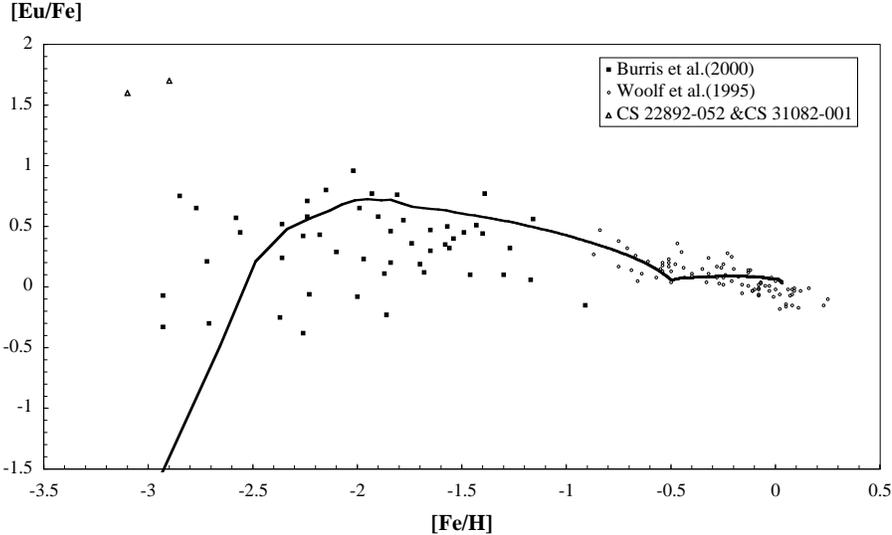}

\caption{The predicted temporal evolution of r-process elements compared to the observed temporal evolution of
Europium in our Galaxy; from De Donder \& Vanbeveren (2003)}
\label{fig:fig6}
\end{center}
\end{figure}

\section{The effects of binaries on the CEM of galaxies}

How binaries can be implemented in a CEM has been described in De Donder \& Vanbeveren (2002, 2003a). In the latter papers
we studied the effects of binaries and massive BH formation on the temporal evolution of $^4$He, $^{12}$C, $^{14}$N,
$^{16}$O, $^{20}$Ne, $^{24}$Mg, $^{28}$Si, $^{32}$S, $^{40}$Ca and $^{56}$Fe.  

Cowan et al. (1991) reviewed the possible astrophysical sites where the r-process can happen. Two sites have favourable
physical conditions in order to be a major r-process source: the SN explosion of a massive star (SN II, SN
Ib/c) (Woosley et al., 1994; Takahashi et al., 1994; Hoffman et al., 1997; Qian \& Woosley, 1996; Meyer et al., 1998) and
the NSM (Davies et al., 1994; Janka \& Ruffert, 1996; Baumgarte et al., 1997; Ruffert \& Janka,
1998; Rosswog et al., 1999, 2000, 2001; Freiburghaus et al., 1999a, b).  De Donder \& Vanbeveren (2003c) combined the CEM
with a detailed PNS model for NSMs in order to simulate the temporal evolution of r-process elements. \\

Conclusions:

\begin{itemize}

\item The rate of star formation only marginally depends on whether binaries form or not.
\item The age metallicity relation predicted by a theoretical CEM hardly depends on whether MCBs are included or not.
\item The temporal evolution of $^4$He, $^{12}$C, $^{14}$N, $^{16}$O, $^{20}$Ne, $^{24}$Mg, $^{28}$Si, $^{32}$S,
$^{40}$Ca and $^{56}$Fe depends on whether or not MCBs are included in a CEM, but the effect is in most cases smaller
than a factor of 2. 
\item NSMs are promising sources of r-process elements. We find an overall good agreement with the observed evolutionary
behaviour of the element europium which is a good representative of the r-process elements (see Figure 6). 
\item Our simulations show that case BB RLOF in massive binaries plays a very important role in the Galactic evolution of
NSM events and the corresponding r-process element enrichment.   
\item We propose that halo stars with a large enhancement in r-process elements and a very low iron abundance (the
triangles in Figure 6), may have been polluted with r-process elements by double NS binaries that were kicked out of the
galactic disk at birth and reached the halo at their moment of merging.

\end{itemize}

\noindent Final remark: the main effect on CEM simulations of stellar rotation during the CHB and CHeB phases of
stars, is  the enlargement of the convective core during CHB. Since, on average, the effect of rotation on the
convective core during CHB is similar as the effect of a moderate amount of convective core overshooting, and since
most of the present day CEMs use evolutionary calculations which account for a moderate amount of convective core
overshooting, {\it most of the CEMs accounted for stellar rotation avant la lettre}. Notice however that
rotation may explain the presence of primary nitrogen during the early phases of a galaxy (Meynet and Maeder, 2002).


\begin{thebibliography}{}
\bibitem{} Belkus, H., Van Bever, J., Vanbeveren, D., Van Rensbergen, W., 2003, A\&A, in press.
\bibitem{} Brandt, W.N., Podsiadlowski, P., Sigurdssen, S., 1995, MNRAS, 277, L35.
\bibitem{} Braun, H., Langer, N., 1995, A\&A, 297, 483
\bibitem{} Cannon, R. C., Eggleton, P.P., Zytkow, A. N., Podsiadlowski, P., 1992, ApJ, 386, 206.
\bibitem{} Cowan, J.J., Thielemann, F.-K., Truran, J.W., 1991, Phys. Rep., 208, 267.
\bibitem{} Davies, M., Benz, W., Piran, T., Thielemann, F.-K., 1994, ApJ, 431, 742.
\bibitem{} De Donder, E., Vanbeveren, D., 1998, A\&A, 333, 557.
\bibitem{} De Donder, E., Vanbeveren, D., 2002 , New Astronomy, 7, 55.
\bibitem{} De Donder, E., Vanbeveren, D., 2003a, New Astronomy, in press.
\bibitem{} De Donder, E., Vanbeveren, D., 2003b, New Astronomy, in press.
\bibitem{} De Donder, E., Vanbeveren, D., 2003c, New Astronomy, in press.
\bibitem{} De Donder, E., Vanbeveren, D., Van Bever, J., 1997, A\&A, 318, 812.
\bibitem{} De Loore, C., Vanbeveren, D., 1992, A\&A, 260, 273.
\bibitem{} Freiburghaus, C., Rosswog, S., Thielemann, F.-K., 1999b, ApJ, 525, L121.
\bibitem{} Freiburghaus, C.,Rembges, J., Rauscher, T., et al., 1999a, ApJ, 516, 381.
\bibitem{} Fryer, C.L., Kalogera, V., 2001, ApJ, 554, 548.
\bibitem{} Habets, G.M.H.J., 1986a, A\&A, 165, 95.
\bibitem{} Habets, G.M.H.J., 1986b, A\&A, 167, 61.
\bibitem{} Hamann, W.-R., Koesterke, L., 1998, A\&A., 333, 251.
\bibitem{} Hamann, W.-R., Koesterke, L., Wessolowski, U., 1995, A\&A, 299, 151.
\bibitem{} Heger, A., Woosley, S.E., Langer, N., Spruit, H.C., 2003, in {\it Stellar Rotation}, eds. A. Maeder and P. Eenens, in press.
\bibitem{} Hillier, D.J., 1996, in {\it Wolf-Rayet stars in the Framework of Stellar Evolution}, eds. J.M. Vreux, A. Detal, D. Fraipont-Caro, E. Gosset, G. Rauw, UniversitŽ de Lige, p. 509.
\bibitem{} Hoffman, R.D., Woosley, S.E., Qian, Y.-Z., 1997, ApJ, 482, 951.
\bibitem{} Humphreys, R.M., Davidson, K., 1994, PASP, 106, 1025.
\bibitem{} Humphreys, R.M., McElroy, D.B.,  1984, ApJ, 284, 565-577.
\bibitem{} Iben, I., Jr., Livio, M., 1993, PASP, 105, 1373.
\bibitem{} Iglesias, C.A., Rogers, F.J., 1996, ApJ 464, 943
\bibitem{} Israelian, G., Garcia, R.J., Rebolo, R., 1998, ApJ, 507, 805.
\bibitem{} Janka, H.T., Ruffert, M., 1996, A\&A, 307, L33.
\bibitem{} Kippenhahn, R., Ruschenplatt, G., Thomas, H.C., 1980, A\&A, 91, 175.
\bibitem{} Kippenhahn, R., Weigert, A., 1967, Z. Astrophys. 65, 251.
\bibitem{} Lauterborn, D.: 1969, in {\it Mass Loss from Stars}, ed. M. Hack, D. Reidel: Publ. Com., Dordrecht p.262.
\bibitem{} Levato, H., Malaroda, S., Morell, N., Garcia, B., Hernandez, C., 1991, ApJ. Supp. 75, 869.
\bibitem{} Lorimer, D.R., Bailes, M., Harrison, P.A., 1997, MNRAS, 289, 592.
\bibitem{} Mason, B.D., Gies, D.R., Hartkopf, W.I., et al., 1998, AJ, 115, 821.
\bibitem{} Mermilliod, J.-C.: 2001, in {\it The Influence of Binaries on Stellar Population Studies}, ed. D. Vanbeveren, Kluwer Academic Publishers: Dordrecht, p. 3.
\bibitem{} Meyer, B.S., McLaughlin, G.C., Fuller, G.M., 1998, Phys. Rev., C, 58, 3696.
\bibitem{} Meynet, G., Maeder, A., 2002, A\&A, 390, 561.
\bibitem{} Moffat, A.F.J., 1996, in {\it Wolf-Rayet stars in the Framework of Stellar Evolution}, eds. J.M. Vreux, A. Detal, D. Fraipont-Caro, E. Gosset, G. Rauw, UniversitŽ de Lige, p. 553.
\bibitem{} Monaghan, J.J., ARA\&A, 30, 543.
\bibitem{} Neo, S., Miyaji, S., Nomoto, K., Sugimoto, D., 1977, PASJ, 29, 249.
\bibitem{} Petrenz, P., Puls, J., 1996, A\&A, 312, 195.
\bibitem{} Petrenz, P., Puls, J., 2000, A\&A, 358, 956..
\bibitem{} Peters, P.C., 1964, Phys. Rev., 136, B1224.
\bibitem{} Qian, Y.-Z., Woosley, S.E., 1996, ApJ, 471, 331.
\bibitem{} Rasio, F.A., Livio, M., ApJ, 471, 366.
\bibitem{} Rasio, F.A., Shapiro, S.L., 1991, ApJ, 377, 559.
\bibitem{} Rasio, F.A., Shapiro, S.L., 1992, ApJ, 401, 226..
\bibitem{} Rosswog, S., Davies, M.B., Thielemann, F.-K., Piran, T., 2000, A\&A, 360, 171.
\bibitem{} Rosswog, S., Freiburgerhaus, C., Thielemann, F.-K., Davies, M.B., 2001, 20th Texas Symposium on relativistic astrophysics, Austin, Texas, 10-15 December 2000, Melville, NY: American Institute of Physics, 2001, xix, 938 p. AIP conference proceedings, Vol. 586, eds; J. Craig Wheeler and Hugo Martel.
\bibitem{} Rosswog, S., Liebendorfer, M., Thielemann, F.-K., Davies, M., Benz, W., Piran, T., 1999, A\&A, 341, 499.
\bibitem{} Schmutz, W., 1996, in {\it Wolf-Rayet stars in the Framework of Stellar Evolution}, eds. J.M. Vreux, A. Detal, D. Fraipont-Caro, E. Gosset, G. Rauw, UniversitŽ de Lige, p. 553.
\bibitem{} Soberman, G. E.; Phinney, E. S.; van den Heuvel, E. P. J., 1997, A\&A 327, 620.
\bibitem{} Sparks, W.M., Stecher, T.P., 1974, ApJ, 188, 149.
\bibitem{} Takahashi, K., Witti, J., Janka, H.-Th, 1994, A\&A, 286, 857.
\bibitem{} Tauris, T. M., Takens, R. J., 1998, A\&A, 330,.1047T
\bibitem{} Thorne, K.S., Zytkow, A.N., 1977, ApJ, 212, 832.
\bibitem{} Van Bever, J., Belkus, H., Vanbeveren, D., Van Rensbergen, W., 1999, New Astronomy, 4, 173.
\bibitem{} Van Bever, J., Vanbeveren, D., 1997, A\&A, 322, 116.
\bibitem{} Van Bever, J., Vanbeveren, D., 1998, A\&A, 334, 21.
\bibitem{} Van Bever, J., Vanbeveren, D., 2000, A\&A, 358, 462.
\bibitem{} Van Bever, J., Vanbeveren, D., 2003, A\&A, in press.
\bibitem{} Vanbeveren, D., 1991, A\&A, 252, 159.
\bibitem{} Vanbeveren, D., 1996, in {\it Evolutionary Processes in Binary Stars}, eds. R. Wijers, M. Davies, C. Tout, Kluwer Academic Publishers: Dordrecht, p. 155.
\bibitem{} Vanbeveren, D., 2001, in {\it The Influence of Binaries on Stellar Population Studies}, ed. D. Vanbeveren,
Kluwer Academic Publishers: Dordrecht, p. 249.
 \bibitem{} Vanbeveren, D., De Donder, E., Van Bever, J., et al.. 1998c, New Astronomy, 3, 443.
\bibitem{} Vanbeveren, D., De Loore, C., 1994, A\&A., 290, 129.
\bibitem{} Vanbeveren, D., Van Rensbergen, W., De Loore, C.: 1998b, The Astron. Astrophys. Review 9, 63.
\bibitem{} Vanbeveren, D., Van Rensbergen, W., De Loore, C.: 1998a, monograph {\it The Brightest Binaries}, eds. Kluwer Academic Publishers: Dordrecht.
\bibitem{} Verschueren, W., David, M., Brown, A. G. A., 1996, in {\it The origins, evolution, and destinies of binary stars in clusters}, Astronomical Society of the Pacific Conference Series, Volume 90, p.131.
\bibitem{} Webbink, R.F., 1984, ApJ, 277, 355. 
\bibitem{} Woosley, S.E., Wilson, J.R., Mathews, G.J., et al., 1994, ApJ, 433, 229.

\end{thebibliography}
\end{document}